\documentclass[doublecol]{epl2}
\input epsf.sty
\usepackage{amsmath}
\usepackage{amssymb}
\usepackage{graphicx}
\usepackage{subfigure}

\title{Cyclic competition of four species: mean field
theory and stochastic evolution}
\shorttitle{Cyclic competition of four species}
\author{Sara O. Case, Clinton H. Durney, Michel Pleimling, and R.K.P. Zia}
\shortauthor{Sara O. Case \etal}
\institute{Department of Physics, Virginia Polytechnic Institute and State University, Blacksburg, VA 24061-0435 USA}

\abstract{Generalizing the cyclically competing three-species model 
(often referred to as the rock-paper-scissors game), we consider a simple 
system of population dynamics without spatial structures that involves four
species. Unlike the previous model, the four form alliance pairs which resemble
partnership in the game of Bridge. 
In a finite system with discrete stochastic dynamics,
all but 4 of the absorbing states consist of coexistence of a partner-pair.
{}From a master equation, we derive a set of mean field equations of
evolution. This approach predicts complex time dependence of the system and
that the surviving partner-pair is the one with the larger product of their
strengths (rates of consumption). Simulations typically confirm these
scenarios. Beyond that, much richer behavior is revealed, including
complicated extinction probabilities and non-trivial distributions of the
population ratio in the surviving pair. These discoveries naturally raise a
number of intriguing questions, which in turn suggests a variety of future
avenues of research, especially for more realistic models of multispecies
competition in nature.}

\begin{document}
\maketitle

\section{Introduction}

Over the years evolutionary game theory and population
dynamics have yielded important insights into biodiversity and the behavior
of multispecies ecological systems \cite{Hof98,Now06,Sza07}. The complexity
of real-world systems makes a full understanding of their properties very
difficult. For that reason, the study of simple model systems is extremely
valuable, as the complete knowledge of these systems allows to identify
generic features valid in the more realistic but also more complex
situations.

In this context multispecies models with cyclic competition constitute some
of the simplest cases where coexistence and species extinction can be
studied using techniques from statistical mechanics and from non-linear
dynamics \cite{Sza07,Fre09}. Many recent investigations revealed a rich and
complex behavior. In particular, for systems with three species (a.k.a.
rock-paper-scissors game) \cite
{Fra96a,Fra96b,Pro99,Tse01,Ker02,Kir04,Rei06,Rei07,Rei08,Cla08,Pel08,Rei08a,Ber09,Ven10,Shi10,And10,Rul10,Wan10,Mob10,He10,Win10}%
, the results range from surprising survival/extinction probabilities in
models with no spatial structure to pattern formation and mobility effects
in one- and two-dimensional lattices. By contrast, far less is known for
systems with more than three species \footnote{%
In lattice models one sometimes speaks of a four-state rock-paper-scissors
games when empty sites are allowed \cite{Pel08,He10}. In the following we do
not consider empty sites to form an independent species.}. Frachebourg 
\textit{et al.} \cite{Fra96a,Fra96b,Fra98} considered $M$ species in
one-dimension, $X_m+X_{m+1}\stackrel{k_m}{\rightarrow }2X_m$ ($m=1,\cdots
,M; $ $X_{M+1}=X_1$), competing with equal rates, $k_m$. The steady states
consist of single-species domains for $M=3,4$, but are qualitatively
different for $M\ge 5$. For systems on two-dimensions with slightly more
complex rates, the segregation process and the formation of defensive
alliances have also been studied \cite{Kob97,Sat02,Sza04,He05,Sza07b,Sza08}.
In a recent paper \cite{Dob10} the Fokker-Planck equations for conserved
quantities where derived for the $M=3$ and $M=4$ cases with equal competition rates.
Motivated by real-world systems, some studies also focused on a large number of
competing species with complicated interaction schemes \cite{Sil92,Sza01}.

In this paper we investigate systematically the properties of a non-spatial
game involving four cyclically competing species with arbitrary rates. Not
surprisingly, the behavior of the system is much richer than the three
species case \cite{Rei06,Ber09}. For example, the number of absorbing states
is not fixed at three (or four) but is $2\left( N+1\right) $, where $N$ is the
number of individuals in the system. Such a result can be intuitively
understood, much like in the game of Bridge, where the four players form
partnerships. As a result, the composition of the end state tends to be 
\emph{coexistence} of partner-species, e.g., $L$ of $X_1$ and $N-L$ of $X_3$%
. A mean field approximation is formulated and studied analytically.
Capturing most of the complicated time evolution of the full, stochastic
model, it describes much of the rich behavior. Of course, it cannot predict
extinction events. To explore those processes, we rely on computer
simulations and discover very complex extinction scenarios that depend on
both the rates and the initial conditions. Unlike the three species case,
our system does not support 'the survival of weakest' or the `law of
stay-out' \cite{Ber09,Tai89,Fre01}. Instead, the best maxim seems to be: ``The prey of
the prey of the weakest is the \emph{least likely} to survive.'' This result
is intuitively reasonable, since the prey of the weakest survives easily
and, in turn, causes \emph{its prey }to die quickly. Notably, this maxim
also applies to the three-species case, as illustrated by say, $X_1$ being the
weakest. Then, `the prey of the prey of the weakest' is $X_3$, the demise of
which is excellent news for $X_1$!

In the next section, we specify our model, discuss its absorbing states, and
provide analytic results in a mean field approach. Much of the system's
qualitative behavior can be understood. However, predicting extinction
probabilities is much more challenging and numerical simulations for
exploring them are discussed in the following section. We end with a summary
and some outlook for future research.

\section{Model specifications and mean field theory}

Our system consists of $N$ individuals each of which is identified as one of four interacting species: $A$, $B$, $C$, and $D$. Endowing the species with cyclic
competition, our dynamics consists of picking a random pair and letting the
interactions 
\begin{eqnarray*}
&&A+B\stackrel{p_a}{\rightarrow }A+A;\,\,B+C\stackrel{p_b}{\rightarrow }B+B
\\
&&C+D\stackrel{p_c}{\rightarrow }C+C;\,\,D+A\stackrel{p_d}{\rightarrow }D+D
\end{eqnarray*}
occur with probabilities $p_m$, $m=a,b,c,d$. Note that $AC$ and $BD$ pairs
are \emph{non-interacting}. Denoting the numbers of each species in our
system by $N_m$, a configuration of the system, which has no spatial
structure, is completely specified by these integers. With $N=\sum_mN_m$
being a constant, our configuration space is actually a set of points within
a regular tetrahedron \cite{Spe80}. Unlike the cyclic competition of three species, we
have $2\left( N+1\right) $ absorbing states here. They form two fixed lines, 
$N_a+N_c=N$ and $N_b+N_d=N$, and describe \emph{coexistence} of the
non-interacting pairs, $A$-$C$ and $B$-$D$, respectively. Moreover, note
that each face of the tetrahedron is also `absorbing,' in the sense that
transitions into the face are irreversible. Within each face, the problem is
a special limit of the three species model, namely, one of the three rates
being zero.

{}From these dynamic rules, it is simple to write a master equation for $%
P\left( \left\{ N_m\right\} ;t\right) $, the probability for finding the
system $t$ steps after an initital configuration $\left\{ N_{m0}\right\} $.
To find its solution is far less simple, however. Instead, we will exploit a
mean field approximation for the evolution of the averages of the \emph{%
fractions}, $A\left( t\right) \equiv \sum_{\left\{ N_m\right\} }\left(
N_a/N\right) P\left( \left\{ N_m\right\} ;t\right) \,$, etc. Following
standard routes, we start from the master equation for $P$ and consider the
large $N$ behavior to arrive at\cite{CDPZ-long} 
\begin{eqnarray}
\partial _tA &=&\left[ k_aB-k_dD\right] A;\,\,\partial _tB=\left[
k_bC-k_aA\right] B  \label{ABCD eqns 1} \\
\partial _tC &=&\left[ k_cD-k_bB\right] C;\,\,\partial _tD=\left[
k_dA-k_cC\right] D  \label{ABCD eqns 2}
\end{eqnarray}
where $A\left( t\right) $ is simplified to $A$, etc. Here, $t$ is regarded
as a continuous variable and the ``rates'' $k_m$ can be related to the
discrete time step, the $p$'s above, and $N$\cite{CDPZ-long}. Of course, the
conservation law now reads $A+B+C+D=1$. Since an overall scale can be
absorbed into $t$, we will follow the normalization in the literature: $%
k_a+k_b+k_c+k_d=1$. The remainder of this section will be devoted to a study
of the evolution of $A\left( t\right) $, $B\left( t\right) $, etc. starting with $%
A\left( 0\right) =A_0\equiv N_{a0}/N$, etc.

Exploiting the exponential nature of typical growth/decay, we write the
above equations as 
\begin{eqnarray}
\partial _t\ln A &=&k_aB-k_dD;\,\,\partial _t\ln C=k_cD-k_bB  \label{lnAC} \\
\partial _t\ln B &=&k_bC-k_aA;\,\,\partial _t\ln D=k_dA-k_cC  \label{lnBD}
\end{eqnarray}
These clearly expose the alliance into opposing pairs $AC$ and $BD$, an
essential feature absent in the three species model. Borrowing the language
of Bridge, we will refer to $AC$ and $BD$ as partner-pairs, as each player
works in favor of its partner and against the opposing pair. To be
quantitative, we construct appropriate linear combinations such as $\partial
_t\left[ k_b\ln A+k_a\ln C\right] =\lambda D$, with 
\begin{equation}
\lambda \equiv k_ak_c-k_bk_d  \label{lambda-def}
\end{equation}
being a crucial parameter. In addition to controlling how each species
affects the growth/decay of the opposing pair, $\lambda $ generates a simple
evolution 
\begin{equation}
Q\left( t\right) =Q\left( 0\right) e^{\lambda t}
\end{equation}
for the quantity 
\begin{equation}
Q\equiv \frac{A^{k_b+k_c}C^{k_d+k_a}}{B^{k_c+k_d}D^{k_a+k_b}}\,\,.
\label{Q-def}
\end{equation}
Similar to $R\equiv A^{k_b}B^{k_c}C^{k_a}$ in the three species system\cite
{Ber09}, $Q$ is $t$-dependent as opposed to $R$ being \emph{invariant}.
Furthermore, since $A,B,C,D$ are bounded by unity, the indefinite
decay/growth in $Q$ can only occur when $A,C$ or $B,D$ vanish. As a result,
the sign of $\lambda $ controls which pair survives. Intuitively, this
prediction seems understandable: The pair with the larger \emph{rate-product}
($k_ak_c$ or $k_bk_d$) wins.

\begin{figure}
\subfigure[]{\includegraphics[width=8.0cm]{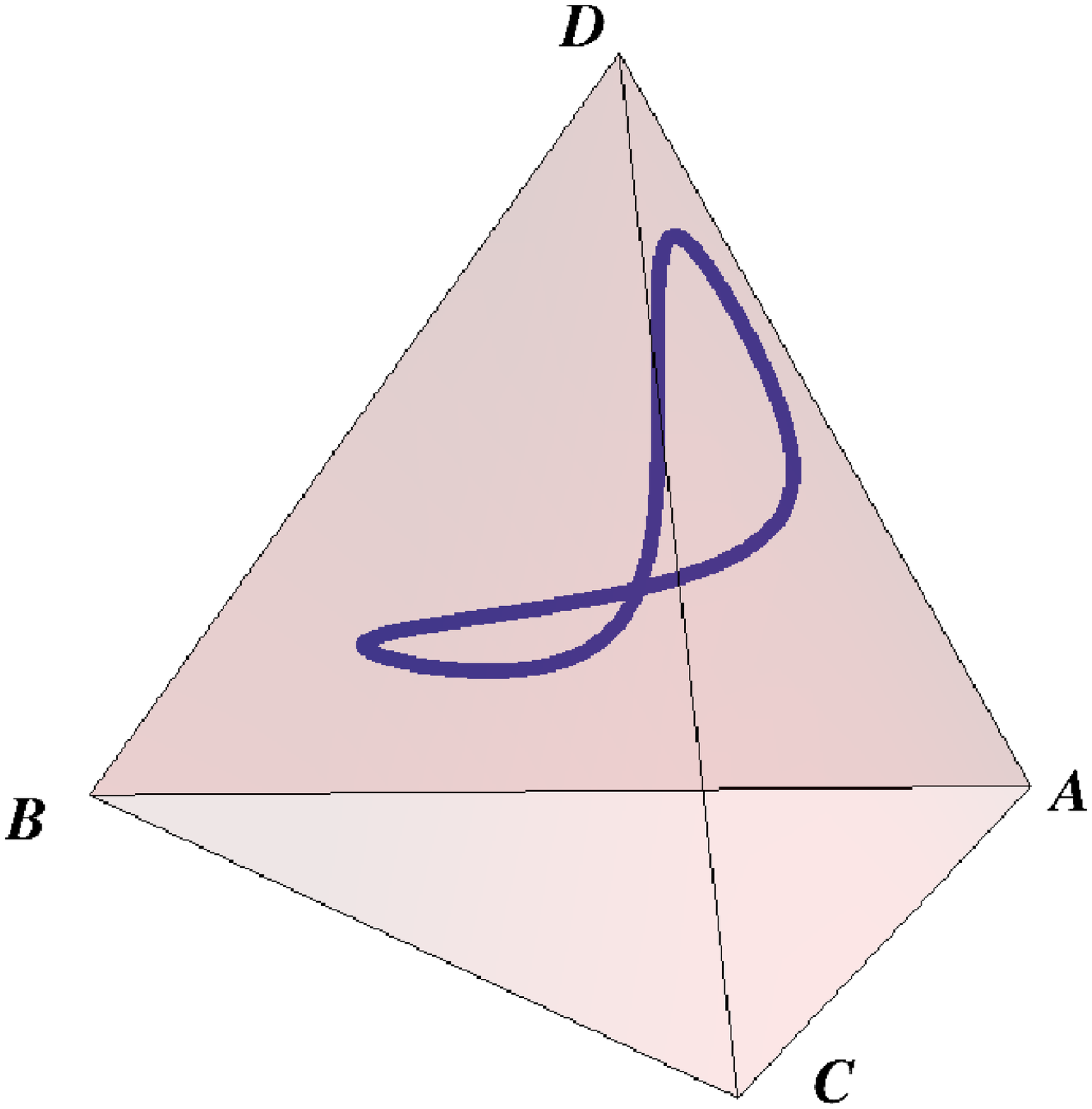}}
\subfigure[]{\includegraphics[width=7.2cm]{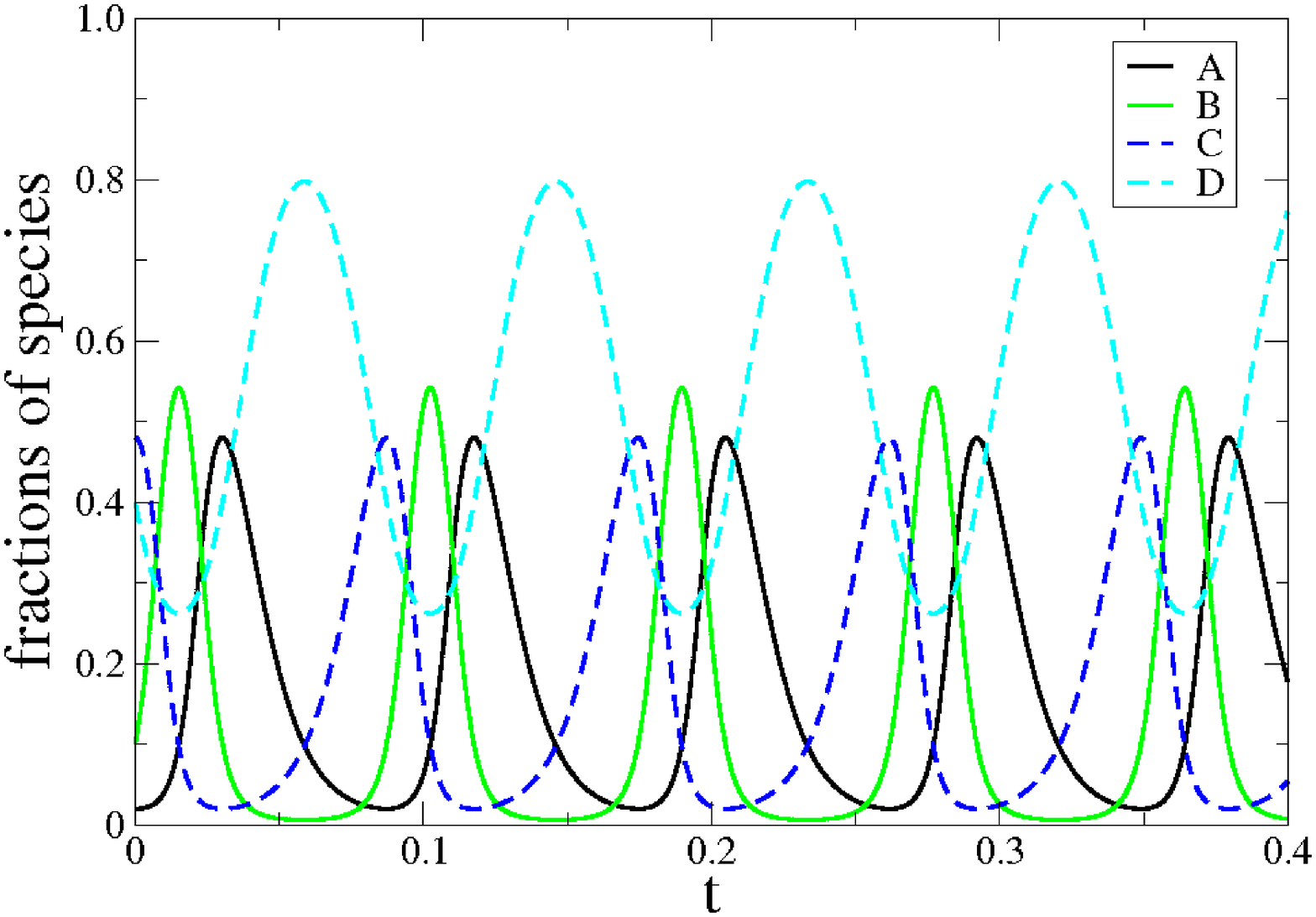}}
\caption{An example of mean field evolution for $\lambda =0$: $\left\{ k_m\right\}
=\left( 0.4,0.4,0.1,0.1\right) $. (a) Closed loop in the tetrahedron forming
the configuration space and (b) the fractions of the different species as a
function of time. These data were generated by applying a fourth order
Runge-Kutta scheme with $\Delta t=10^{-5}$ on Eqns. (\ref{ABCD eqns 1},\ref
{ABCD eqns 2}).
        }
\label{fig1}
\end{figure}

Obviously, systems with $\lambda =0$ are special, as $Q$ is a constant of
the motion. Indeed, there are \emph{two} invariants, which can be simply $%
A^{k_b}C^{k_a}$ and $B^{k_d}D^{k_a}$ (as generalizations of $AC$ and $BD$ in 
\cite{Sza07,Dob10}, where $k_m=1,\,\,\forall m$). Fixing these by the initial
conditions ($A_0,B_0,C_0,D_0$), we define natural variables: $\rho _A\equiv
\left( A/A_0\right) ^{k_b}$, etc., which obey 
\begin{equation}
\rho _A\rho _C=1=\rho _B\rho _D.
\end{equation}
These equations define hyperbolic sheets through the tetrahedron and their
intersection is a closed loop that resembles (the rim of) a saddle. Fig. \ref
{fig1}a shows an example of such an orbit, for the case $k_a=k_b=0.4$ and $%
k_c=k_d=0.1$. Meanwhile, Fig. \ref{fig1}b shows the associated ever-lasting
oscillations in $A,B,C,D$. 

\noindent One important characteristic of a closed orbit is its extremal
points. For example, let $A_{\pm }$ denote the largest/smallest values $A$
assumes, given an initial point ($A_0,B_0,C_0,D_0$) and a set of $k$'s.
Then, $A_{\pm }$ are solutions to a generically transcendental equation: $%
A_0^{-k_b/k_a}A_{\pm }+C_0A_{\pm }^{-k_b/k_a}=$ constant, which depends on $%
B_0,D_0,$ and the $k$'s. Typically, two distinct solutions exist,
corresponding to the two extremes. At these points, the values assumed by $%
B,C,D$ are, in general, not extremal themselves. While $C=C_0\left(
A_0/A_{\pm }\right) ^{k_b/k_a}$ at these points, $B$ takes on the same value
at both turning points: $\left( k_d^{k_a}k_a^{-k_a}B_0^{k_d}D_0^{k_a}\right)
^{1/\left( k_a+k_d\right) }$. Similarly, $D$ is also unique. When $%
A_{+}=A_{-}$, we are at a \emph{fixed line} -- formed by the intersection of
the two planes: $k_aA=k_bC$ and $k_aB=k_dD$. Unlike the lines of absorbing
states ($A$-$C$ and $B$-$D$), points on this line are neither stable nor
stationary under the stochastic dynamics. Being straight and bridging the $A$%
-$C$, $B$-$D$ lines, this fixed line is enclosed by every closed orbit. In
its neighborhood, these orbits approach circles, on which the system `moves'
with $\omega \propto $ $\sqrt{k_ak_c}$. Details supporting these remarks
will be provided in a future publication\cite{CDPZ-long}.

\begin{figure}
\subfigure[]{\includegraphics[width=8.0cm]{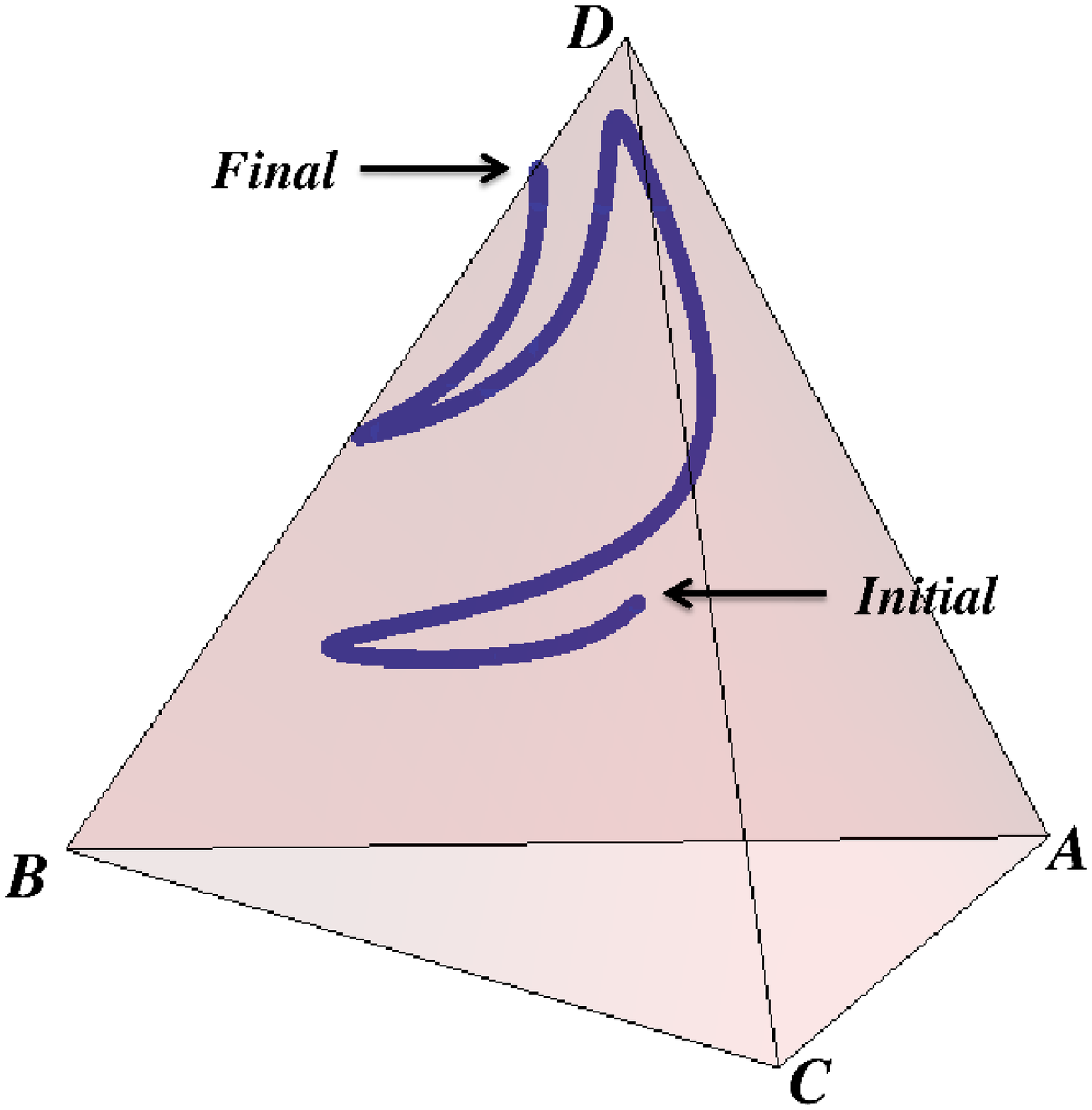}}
\subfigure[]{\includegraphics[width=8.0cm]{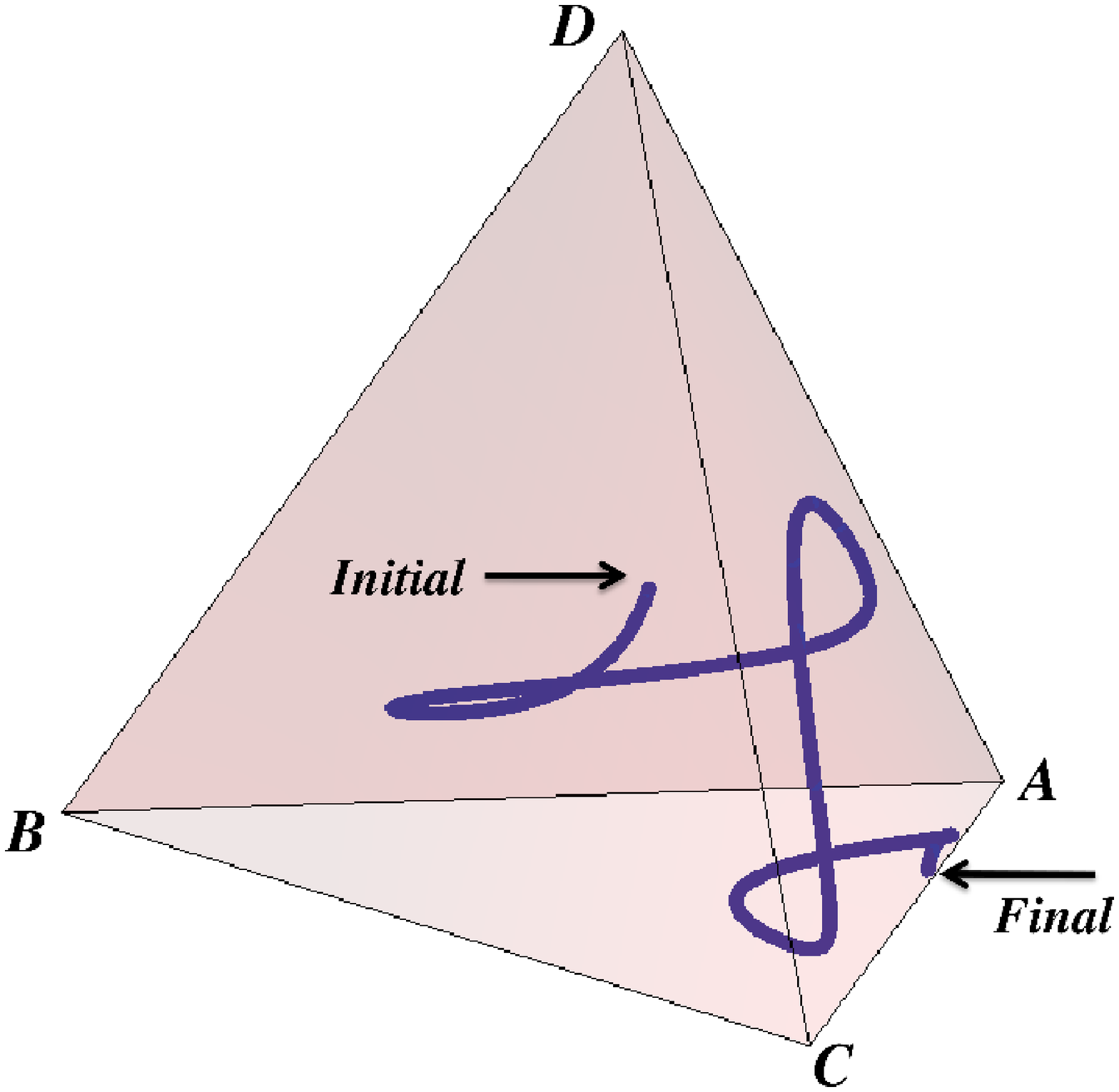}}
\caption{Examples of mean field evolution for $\lambda \neq 0$: (a) $\left\{
k_m\right\} =\left( 0.45,0.33,0.14,0.08\right) $ with $\lambda =.0366$, and
(b) $\left\{ k_m\right\} =\left( 0.35,0.42,0.09,0.14\right) $ with $\lambda
=-.0273$. The initial fractions in both are $\left\{ N_{m0}/N\right\} =\left(
0.02,0.10,0.48,0.40\right) $ Note the two trajectories end on the $A$-$C$
and $B$-$D$ lines, respectively. These data were generated by the same scheme
as in Fig. \ref{fig1}.
        }
\label{fig2}
\end{figure}

For systems with $\lambda \neq 0$, non-trivial fixed points cannot exist (as 
$\ln Q\rightarrow \lambda t$). Thus, in a finite system, extinction of one
of the species must occur quite rapidly. Fig. \ref{fig2} shows two typical
cases, one for each sign of $\lambda $. Although the mean field provides
good fits for short times, it predicts neither the (average) time for the
first species to die out nor the composition of the other three at this
extinction event. Nevertheless, we can again rely on mean field theory \emph{%
after} the system `lands' on an absorbing face (of the tetrahedron). For
example, if $D$ vanishes first and the system consists of ($A_i,B_i,C_i,0$)
at that time, then mean field theory predicts the system to end at a point
on the $A$-$C$ line: ($A_f,0,1-A_f,0$). Here $A_f$ is the larger of the two
solutions to another transcendental equation: $A_f^{k_b}\left( 1-A_f\right)
^{k_a}=A_i^{k_b}C_i^{k_a}$. As will be shown below, these predictions are
born out quite well in finite, stochastic systems.

\section{Stochastic evolution: exact and simulation results} 

For a full 
stochastic process, there are limitations to a mean field approach. In
particular, it fails when any $N_m$ is not `large,' e.g., in small systems,
or near extinction events. If a system is `very small,' numerically exact methods
can be exploited to find exact solutions to the master equation. Indeed, for
the smallest, non-trivial system ($N=4$), simple algebra is sufficient for
finding analytic expressions for \emph{all} transition probabilities (from
any state to any other) and for \emph{arbitrary }rates. Unlike $N=3$ in
three-species\cite{Ber09} however, the results are not trivially linear (in 
$k_m$). Deferring details to elsewhere \cite{CDPZ-long}, we only present some
general observations here. Though there is a finite probability that the
`weakest' is the lone survivor, there is also a good chance for (one or both
of) its \emph{opponent pair} to survive. The clearest conclusion is: When
the consumption rate of the weakest approaches zero, the survival probability of
its \emph{partner} vanishes. These considerations led us to a general maxim:
``The prey of the prey of the weakest is the least likely to survive.'' As
pointed out above, this maxim is consistent with ``survival of the weakest''
in 3-species models.

For systems with larger $N$'s (say, $\gtrsim 100$), Monte Carlo techniques
are necessary to uncover interesting behaviors in our system. To speed up
the runs, we exploit the Gillespie updating scheme, in which an interaction
always occur at each `step' (with appropriate \emph{relative}
probabilities) \cite{BKL,Gil}. By contrast, the
standard scheme is much slower, as many randomly chosen pairs do not
interact. Of course, the detailed $t$ dependence will be quite different,
so that direct comparisons with mean field predictions are not possible.
Nevertheless, we can rely on this scheme for drawing conclusions on survival
and coexistence (as we can show that the extinction probabilities are
scheme-independent). In particular, starting with many random initial
conditions (typically 20000) and a variety of rates, simulations with $N=100~K$ confirm the
mean field predictions, namely, runs for $\lambda \neq 0$ systems ending
with the correct partner pairs and runs for $\lambda =0$ cases failing to
end \cite{TJ}. In the remainder
of this letter, we will focus on a few particular systems with intermediate $%
N$'s, thereby emphasizing extinction processes and highlighting the
differences between stochastic and mean field trajectories.

\begin{figure}
\subfigure[]{\includegraphics[width=8.0cm]{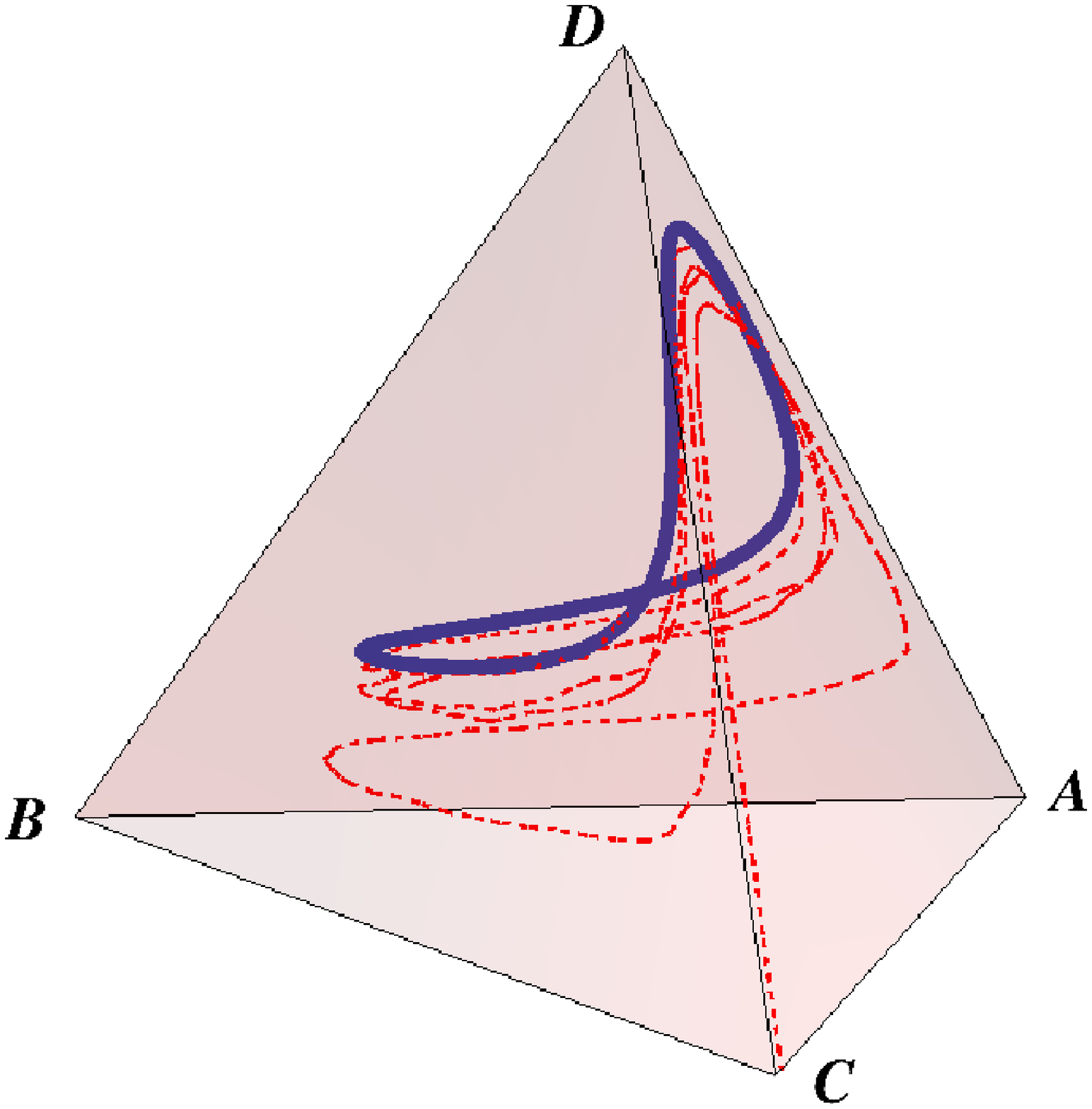}}
\subfigure[]{\includegraphics[width=8.0cm]{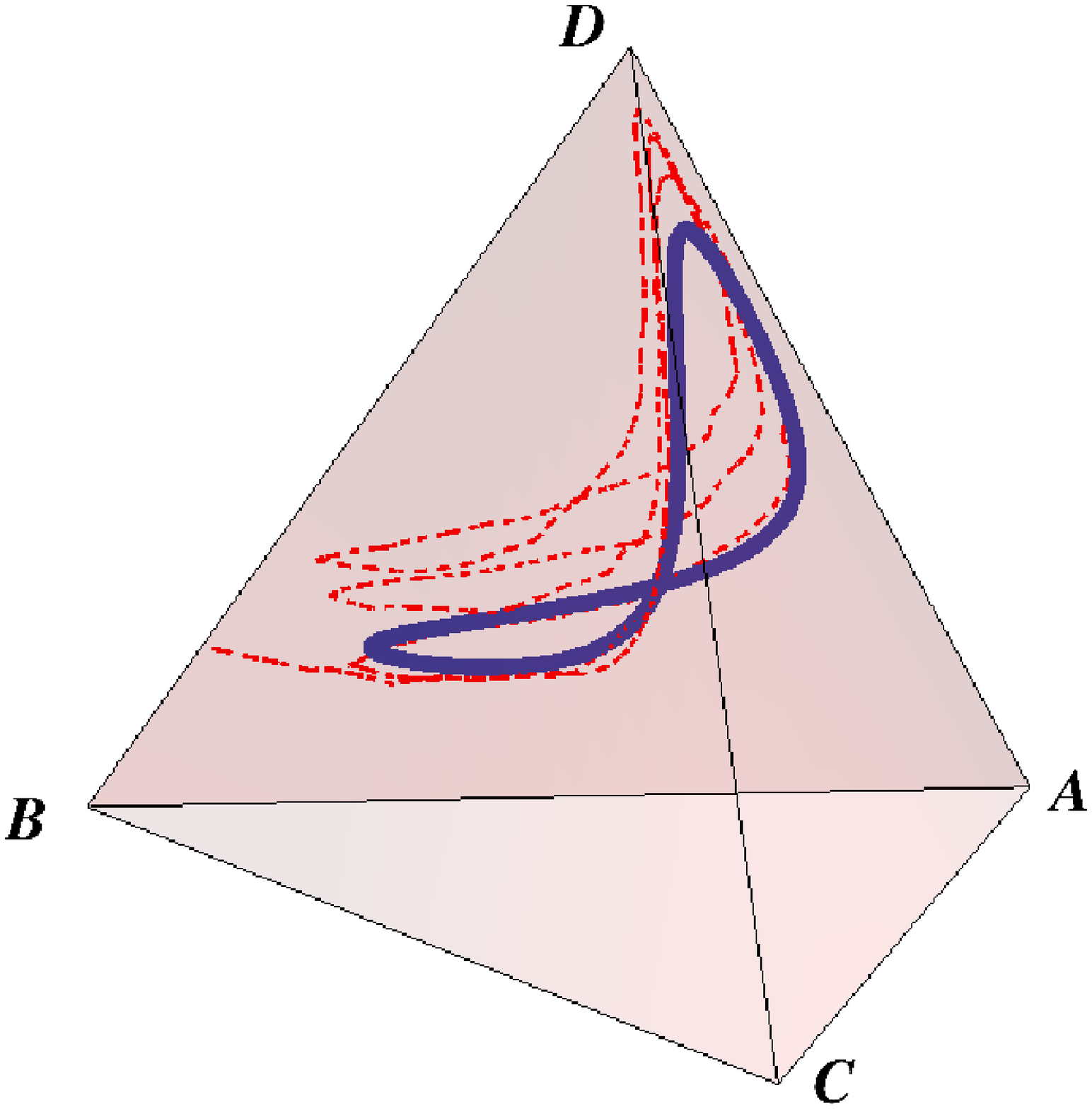}}
\caption{Two examples of stochastic evolution (thin dashed lines, red online) of the
system in Fig. \ref{fig1}, both with $\left\{ N_{m0}\right\} =\left(
0.02,0.10,0.48,0.40\right)\times 1000 $. They follow the mean field loop (thick line,
blue online) initially, but diverge eventually, ending with opposite
partner-pairs: (a) on the $A$-$C$ line and (b) on the $B$-$D$ line. 
        }
\label{fig3}
\end{figure}

As expected, this difference is most pronounced for $\lambda =0$ systems.
Instead of closed orbits (blue online in figures), the trajectories in a
finite stochastic system (red online) end in an absorbing state. In the
3-species case, they mostly end with the `weakest' species as sole
survivors. In our model, no simple conclusions can be drawn. Figs. \ref{fig3}%
a,b illustrate a good example, where runs with identical initial conditions
end very differently. Specifically, we have $N=1K$, initial fractions ($0.02$%
, $0.10$, $0.48$, $0.40$) and rates ($0.4$, $0.4$, $0.1$, $0.1$). The
stochastic trajectories follow these closed loops closely at early times.
But, the noise drives $Q$ away from $Q\left( 0\right) $, so that they later
diverge significantly and, after one of the four species dies out, end
rapidly on an absorbing state. Though both runs have the \emph{same} 
$\left\{ N_{m0},k_m\right\} $, the final states in Fig. \ref{fig3}a and b
consist of \emph{opposite} partner-pairs: $AC$ and $BD$, respectively. In
addition, there are non-trivial distributions of survival fractions within
each pair (e.g., $A_f$, when $AC$ survives). Further details will be
published elsewhere \cite{CDPZ-long}.

\begin{figure}
\subfigure[]{\includegraphics[width=8.8cm]{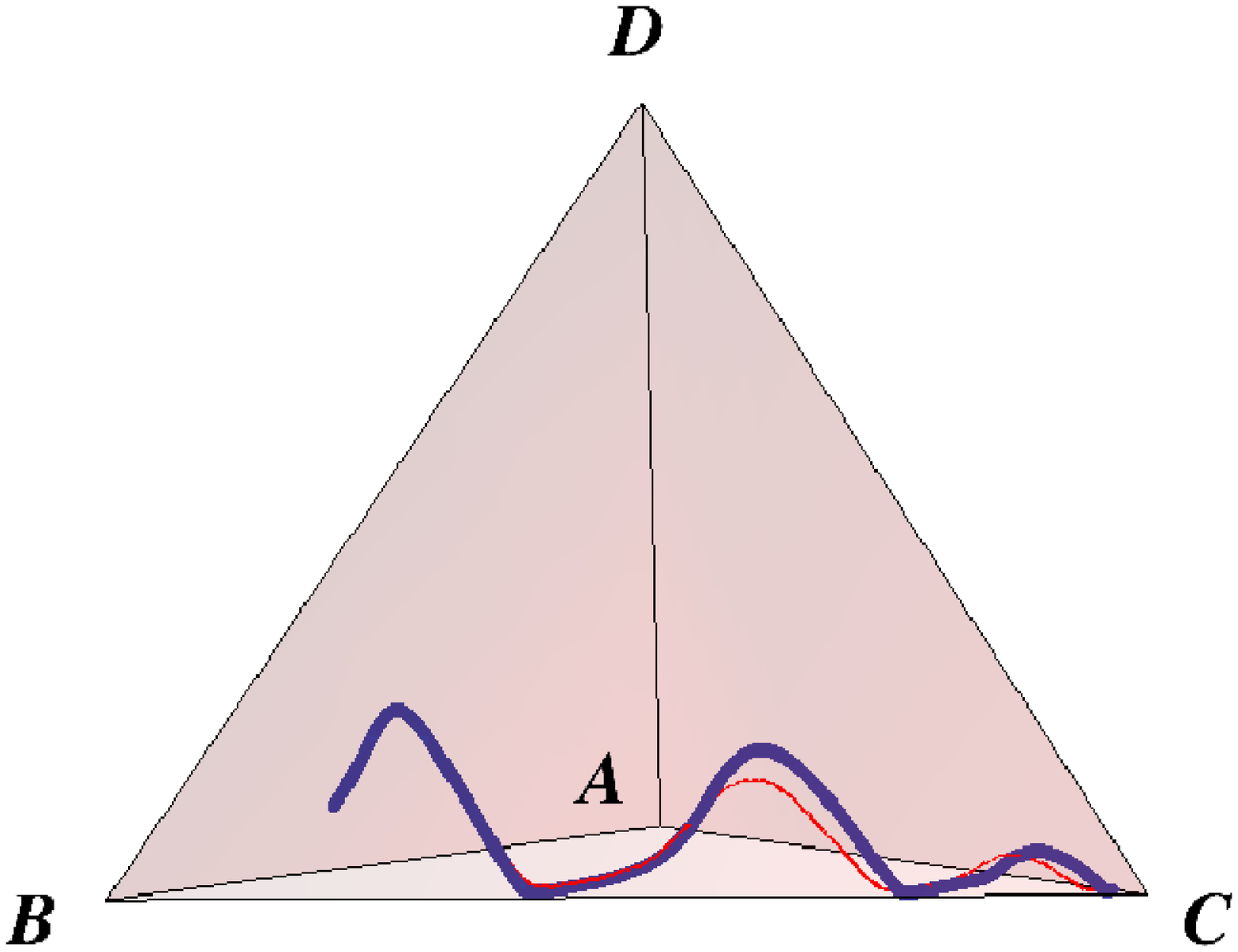}}
\subfigure[]{\includegraphics[width=7.2cm]{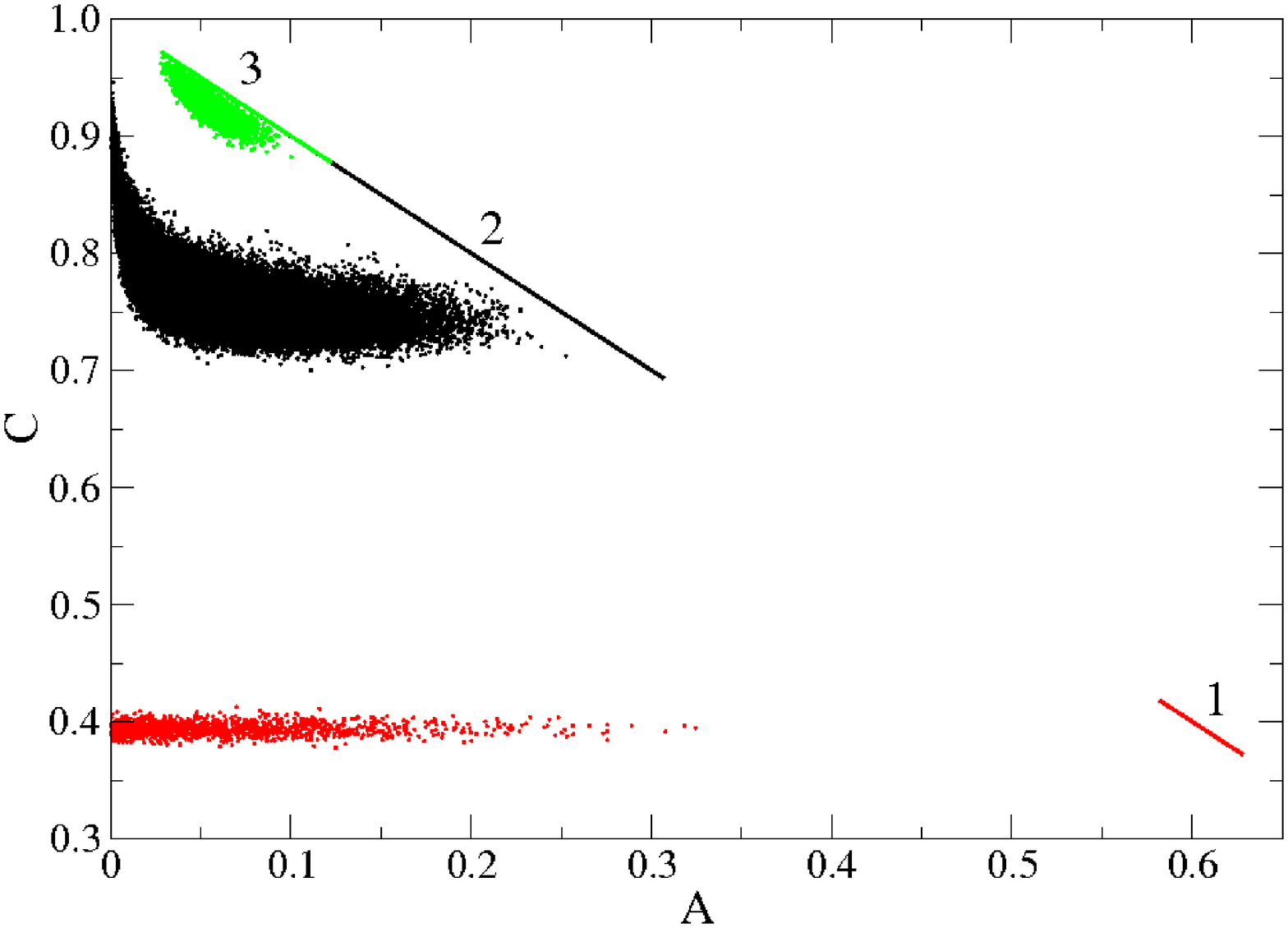}}
\caption{A system with `extreme' rates, $\left\{ k_m\right\} =\left(
0.1,0.0001,0.1,0.7999\right) $, starting with $\left\{ N_{m0}\right\}
=\left( 0.1,0.7,0.1,0.1\right) $. (a) One stochastic trajectory (thin dashed
line, red online) and the mean field evolution (thick line, blue online).
(b) Scatter plot of the composition in 8870 runs of the system at the moment 
$D$ becomes extinct (i.e., `landing sites' on the $ABC$ face), shown as
three separate clusters (red, black, green online) with $A+C<1$. Each run
ends on an absorbing state (a point on the $A+C=1$ line). Also shown is the
scatter plot of these associated states, with the three clusters labeled,
respectively, by 1,2, and 3.
        }
\label{fig4}
\end{figure}

Systems with $\lambda \neq 0$ may appear uninteresting, as they evolve
quickly towards absorbing states. However, we discover rather complex
behavior, especially for systems with \emph{extreme} rates. Let us provide
one illustration, with $N=1K$, initial fractions ($0.1$, $0.7$, $0.1$, $0.1$%
), rates ($0.1$, $0.0001$, $0.1$, $0.7999$) and $10K$ independent trails.
Since $\lambda >0$, the survival rate of the weakest ($B$) is low ($%
\sim 10\%$), while 90\% of the runs end on the $AC$ line. Fig. \ref
{fig4}a shows one particular stochastic trajectory (red online) in the
latter class, as well as the mean field orbit (blue online). Note that they
come very close to the $ABC$ face, (i.e., $D\ll 1$) in two earlier
occasions. Not surprisingly, in these close encounters, many runs actually
`land' on this face. Fig. \ref{fig4}b displays the `landing sites' ($A_i,C_i$
with $B_i=1-A_i-C_i$) from these runs. Notably, they fall into three
clusters (red, black, green online). Predicting the remarkable shape of the
black cluster will undoubtedly be a serious challenge! From here, the system
quickly evolves to the $A$-$C$ line, into similarly colored clusters.
Associated with the unusual black cluster shape, we find the distribution of
the final $A_f$ to be highly skewed and non-Gaussian. More details, as well
as possible explanations, will be provided in a later publication. Here, let
us focus on the evolution from a `landing site' ($A_i,C_i$) to the final
point ($A_f,C_f$). For the $\sim 9K$ runs in this series, we compute 
$\kappa = \ln \left( A_f/A_i\right) +1000\ln \left( C_f/C_i\right) $ as a sensitive
test of the mean field prediction that $\left( A_f/A_i\right) ^{k_b}\left(
C_f/C_i\right) ^{k_a}=1$. Though not identically zero, less than 4\% of the values 
of $\kappa$ (out of the 
$\sim 9K$ values) are outside the range $\left[ -1,1\right] $! It is fair to
conclude that, in this respect, the mean field approach is extremely
successful.

\section{Summary and outlook}

In this article, we investigate the time dependent behavior and extinction
probabilities of a simple model of population dynamics: $N$ individuals of
four different `species,' competing cyclically. Though seemingly a trivial
extension from a similar three-species game (rock-paper-scissors), this
system displays much richer phenomena. Since the configuration space here is
(the interior of) a tetrahedron rather than a triangle, trajectories of the
system may twist and turn in 3-d. Since the four form `partner pairs', much
like in the game of Bridge, the end states typically consists of one of the
pairs, with $N-1$ possible compositions in each case. As a result, there are 
$2\left( N+1\right) $ absorbing states (instead of just 3 or 4), with
generally non-trivial distributions among them. The faces of the tetrahedron
are also `absorbing' (in that they correspond to the extinction of at least
one species); yet the trajectories on them are not just trivial straight
lines. Concerning extinction scenarios, a law gleaned from previous studies
-- `survival of the weakest' -- seems to be violated here. Instead, our
observations support, most consistently, a different maxim: ``The prey of
the prey of the weakest is the \emph{least likely} to survive.'' Much easier
to understand at the intuitive level, this maxim also applies to the
three-species game, where the demise of the prey of one's prey also enhances
one's survival!

Using a mean field approach and computer simulations, we report a number of
other notable findings. Similar to, but more interesting than, $R$ in \cite
{Ber09}, our quantity $Q$ (Eqn. \ref{Q-def}) grows/decays exponentially in
mean field theory and serves as an excellent indicator for which partner
pair will survive. This theory is also quite successful in predicting the
evolution of the stochastic system, as long as (a) no species is close to
extinction, \emph{or} (b) one species is extinct. These two seemingly
contradictory conditions can be easily reconciled, once we are reminded that
mean field theories do not account for discrete variables (zero not being a
`variable' in our dynamics!). With simulations, we discovered complex
extinction scenarios (e.g., non-trivial clustering in Fig. \ref{fig4}b)
displayed by the stochastic model. However, to predict the properties of
such distributions will be a serious challenge, as will be the task for
computing the essentials of  $P\left( \left\{ N_m\right\} ;t\right) $. Work
is in progress to study a related, simpler problem: How does the
distributions of $R$'s and $Q$'s evolve? 

Clearly, the scope of our study is quite limited, so that many natural
questions can be raised. Strictly cyclic competition in multiple species is
rare in reality. What other surprises can we expect if we incorporate into
our four species some of the other \emph{twelve} possible rates? Similarly,
if we introduce realistic birth/death rates (for biological species, e.g.),
will the lack of $N$ conservation produce novel behavior? Needless to say,
the possibilities for generalization (e.g., to $M>4$ species) are limitless,
even for system with no spatial structure. In fact, effects similar to those
discussed here are expected for other even numbers of species \cite{Sat02,Sza08}.
Finally, as we noted in the
Introduction, qualitatively new phenomena (e.g., clustering, pattern
formation, moving fronts) tend to emerge when a population dynamics is placed
on some underlying spatial structure. For example, in spatial systems coexistence
of all four species can be maintained even for $\lambda \ne 0$ \cite{Sza08}. Of course, in nature, different
species are likely to compete in \emph{inhomogeneous} environments. Models
that include such realistic settings \cite{Dob08} will certainly
display a richer variety of properties and will, hopefully, lead to a better
understanding of population dynamics. 

\acknowledgments
We thank the statistical physics group at Virginia Tech for illuminating
discussions and especially T. Jia for communicating his findings on simulations of systems with very large N. This work was supported in part by the US National Science Foundation through Grants DMR-0705152 and DMR-0904999.

\end{document}